# Влияние режима формирования покровного слоя на фотолюминесценцию квантовых точек InAs в кремнии

Вера Вадимовна Лендяшова[1✉], Игорь Владимирович Илькив[2], Вадим Геннадьевич Талалаев[3], Талгат Шугабаев[4], Родион Романович Резник[5], Георгий Эрнстович Цырлин[6,7]

[1, 2, 3, 4, 5, 6]Санкт-Петербургский государственный университет, Санкт-Петербург, Россия

[1, 2, 4, 5, 6]Академический университет им. Ж. И. Алфёрова, Санкт-Петербург, Россия

[6]Институт аналитического приборостроения Российской академии наук, Санкт-Петербург, Россия

[7]Университет ИТМО, Санкт-Петербург, Россия

[1]erilerican@gmail.com        https://orcid.org/0000-0001-8192-7614

[2]fiskerr@ymail.com        https://orcid.org/0000-0001-8968-3626

[3]vadimtal2212@gmail.com        https://orcid.org/0000-0003-2559-889X

[4]talgashugabaev@mail.ru        https://orcid.org/0000-0002-4110-1647

[5]moment92@mail.ru        https://orcid.org/0000-0003-1420-7515

[6]cirlin.beam@mail.ioffe.ru        https://orcid.org/0000-0003-0476-3630

Корреспондирующий автор: Вера Вадимовна Лендяшова, erilerican@gmail.com, +7(921) 776-07-60

**Аннотация**

**Предмет исследования.** Эпитаксиальные слои кремния Si с внедренными квантовыми точками арсенида индия InAs. **Цель работы.** Экспериментальное исследование люминесцентных характеристик квантовых точек InAs в матрице Si и установление зависимости интенсивности фотолюминесценции гетероструктуры от температуры и методики роста покровного слоя кремния. **Метод.** Эпитаксиальные слои Si с внедренными квантовыми точками InAs получены технологией молекулярно-пучковой эпитаксии. Оптические свойства исследовались методом низкотемпературной фотолюминесценции при 10–120 К. **Основные результаты.** Исследовано влияние режимов роста покровного слоя кремния на оптические свойства гетероструктур с субмонослойными квантовыми точками InAs, внедренными в матрицу кремния. Получен сигнал фотолюминесценции от субмонослойных квантовых точек на 1,65 мкм при пониженных температурах вплоть до 120 К. Установлено, что применение двухстадийного способа заращивания InAs наноостровков кремнием позволяет повысить интенсивность фотолюминесценции за счет улучшения кристаллического качества гетероструктур. Анализ температурной зависимости позволил рассчитать энергию активации электронов, удерживаемых в потенциальной яме квантовой точки, сравнимую с термической энергией при комнатной температуре (30 и 25 мэВ соответственно). **Практическая значимость**. Полученные в работе результаты исследования оптических свойств гетероструктур с субмонослойными квантовыми точками InAs могут послужить основой для разработки новых оптоэлектронных приборов на основе кремниевых технологий.

**Ключевые слова:** квантовые точки, арсенид индия, молекулярно-пучковая эпитаксия, полупроводники, кремний, гетероструктуры

# Influence of capping layer growth mode on the photoluminescence of InAs quantum dots in silicon

Vera V. Lendyashova[1✉], Igor V. Ilkiv[2], Vadim G. Talalaev[3], Talgat Shugabaev[4], Rodion R. Reznik[5], George E. Cirlin[6,7]

[1, 2, 3, 4, 5, 6]St. Petersburg State University, St. Petersburg, Russia

[1, 2, 4, 5, 6]Alferov University, St. Petersburg, Russia

[6]Institute for Analytical Instrumentation of the Russian Academy of Sciences, St. Petersburg, Russia

[7]ITMO University, St. Petersburg, Russia

[1]erilerican@gmail.com        https://orcid.org/0000-0001-8192-7614

[2]fiskerr@ymail.com         https://orcid.org/0000-0001-8968-3626

[3]vadimtal2212@gmail.com          https://orcid.org/0000-0003-2559-889X

[4]talgashugabaev@mail.ru         https://orcid.org/0000-0002-4110-1647

[5]moment92@mail.ru          https://orcid.org/0000-0003-1420-7515

[6]cirlin.beam@mail.ioffe.ru          https://orcid.org/0000-0003-0476-3630

## Abstract

**Subject of study.** Silicon Si epitaxial layers with embedded indium arsenide InAs quantum dots. **Aim of study.** Experimental study of the luminescent characteristics of InAs quantum dots in a Si matrix and analyze the relationship between the temperature and growth method of the capping Si layer and the photoluminescence intensity of the heterostructure. **Method.** Si epitaxial layers with embedded InAs quantum dots were obtained by molecular beam epitaxy technology. The optical properties were studied using low-temperature photoluminescence at $10-120$ K. **Main results.** The influence of growth regimes of the silicon capping layer on the optical properties of heterostructures with submonolayer InAs quantum dots embedded in a silicon matrix has been studied. The photoluminescence signal at $1.65$ μm from submonolayer quantum dots at low temperatures up to 120 K was obtained. It was established that the use of a two-stage method of silicon overgrowing InAs nanoislands makes it possible to increase the photoluminescence intensity by improving the crystalline quality of heterostructures. Analysis of the temperature dependence allowed us to calculate an activation energy for electrons confined in the quantum dot's potential well at the level of the thermal energy at room temperature (30 and 25 meV respectively).

**Practical significance.** The results obtained in the study of the optical properties of heterostructures with submonolayer InAs quantum dots can serve as a basis for the development of new optoelectronic devices based on silicon technologies.

**Acknowledgment:** Optical measurements were carried out with the financial support of the Russian Science Foundation grant № 23−79−01117. For growth of experimental samples the authors acknowledge Saint-Petersburg State University for a research project 87465891. Structural properties of grown samples were studied under support of the Ministry of Science and Higher Education of the Russian Federation, research project no. 2019-1442 (project reference number FSER-2020-0013).

**For citation:** Lendyashova V.V., Ilkiv I.V., Talalaev V.G., Shugabaev T., Reznik R.R., Cirlin G.E. Optical properties of InAs quantum dots in a Si matrix grown by molecular beam epitaxy [in Russian] // Opticheskii Zhurnal. 2024. V. XX. № X. P. XX–XX. http://doi.org/XXXX

## ВВЕДЕНИЕ

Интеграция светоизлучающих приборов на основе $A^3B^5$ материалов с кремниевыми структурами представляет значительный интерес в связи с открывающимися возможностями создания новых устройств для фотонных интегральных схем [1,2]. Наряду со сращиванием полупроводниковых пластин [3] повышенный интерес представляет развитие методов монолитной интеграции, т.е. прямого эпитаксиального синтеза гетероструктур на кремниевых подложках. На сегодняшний день значительные успехи достигнуты в получении планарных слоев на основе GaP, который практически согласован с кремнием [4], а также GaAs, являющегося базовым материалом для создания оптических излучателей в ближнем ИК диапазоне [5,6]. Так, например, уже сообщалось о создании лазеров на InAs/GaAs с электрической накачкой и непрерывным режимом работы при комнатной температуре [7], фотоэлектродов из слоев $GaP_{1-x}As_x$ во всем диапазоне составов [8] и двухфункциональных светоизлучающих и фотоприемных гетероструктур на основе GaAsPN [9], выращенных на подложках Si. Однако, несмотря на достигнутые результаты, технология роста буферных слоев с достаточно низкой плотностью дислокаций и отсутствием антифазных доменов до сих пор представляет высокую технологическую сложность [10,11]. В связи с этим является актуальным поиск новых подходов к синтезу гетероструктур, в том числе использование в активной области массивов самоорганизующихся квантовых точек (КТ). Кроме того, уникальные свойства КТ, связанные с релаксацией упругих напряжений, могут быть использованы и для создания принципиально нового типа полупроводниковых гетероструктур – монолитно внедренных в кремниевые матрицы КТ. Формирование таких структур осуществляется путем выращивания наноостровков $A^3B^5$ соединений, например InAs, на поверхности кремния и их последующего эпитаксиального заращивания покровным слоем кремния [12–14]. Несмотря на успехи в их синтезе, оптически активные гетероструктуры, демонстрирующие фотолюминесценцию (ФЛ) в области 1,3 мкм, удалось получить лишь авторам оригинального метода [15,16]. Одной из причин низкого оптического качества гетероструктур может являться высокая дефектность относительно больших КТ InAs, для которых толщина осажденного материала составляла 2–7 монослоев (МС) [14]. Недавно нами было установлено, что формирование КТ при субмонослойных толщинах осажденного InAs (около 0,3–0,7 МС) позволяет улучшить оптическое качество таких гетероструктур и получить сигнал люминесценции в области длин волн 1,6 мкм [17]. При этом значительное влияние на оптические свойства гетероструктур с КТ InAs, внедренными в кремний может оказывать непосредственно кремниевая матрица. Основной целью настоящей работы является изучение влияния режима формирования покровного слоя на оптические свойства квантовых точек InAs, внедренных в кремнии.

## МАТЕРИАЛЫ И МЕТОДЫ

### Гетероструктуры с квантовыми точками InAs

Синтез гетероструктур осуществлялся на установке молекулярно-пучковой эпитаксии (МПЭ) Riber Compact 21 EB200, оснащенной эффузионными источниками для роста соединений $A^3B^5$, а также электронно-лучевыми испарителями для Si и Ge. В качестве подложек использовались Si(100) пластины с разориентацией 4° в направлении [110]. Для удаления слоя окисла и получения гладкой поверхности подложек кремния осуществлялась предварительная химическая обработка по описанному ранее методу [18]. Затем непосредственно в ростовой камере МПЭ установки проводился термический отжиг пластин при 950 °C и формирование буферного слоя кремния толщиной 60 нм при температуре 600 °C. Скорость осаждения Si составляла 0,18–0,27 Å/с. После этого температура подложки понижалась и осуществлялся самоорганизующийся рост наноостровков с эффективной толщиной осажденного слоя InAs равной 0,7 МС (2,1 Å) при температуре подложки 400 °C. Высота сформировавшихся наноостровков составляла 5±2 нм [19]. Скорость осаждения In соответствовала 0,12 Å/с, соотношение потоков материалов V/III было равным 20. На последнем этапе осуществлялось формирование покровного слоя кремния толщиной 30 нм. После выращивания слоя Si образцы выгружались из установки МПЭ для исследования оптических свойств.

### Спектры фотолюминесценции гетероструктур с квантовыми точками InAs

Измерения оптических свойств проводились в He–криостате замкнутого цикла при температурах 10–120 К. Накачка осуществлялась непрерывным лазерным излучением на длине волны 405 нм с использованием полосового фильтра BG-23. Спектры ФЛ регистрировались с помощью InGaAs–фотодетектора (Hamamatsu, GT7754–01) с длиной волны отсечки 2,4 мкм (0,5 эВ). Спектры ФЛ нормировались на спектральную чувствительность фотодетектора. Ширина входной и выходной щелей монохроматора составляла 1–1,5 мм, а время накопления сигнала равнялось 0,5 секундам.

**Структурные свойства гетероструктур с квантовыми точками InAs**

Для исследования размеров и формы квантовых точек, внедренных в матрицу кремния, использовалась установка просвечивающей электронной микроскопии (ПЭМ) Zeiss Libra 200FE при напряжении ускоряющего поля 200 кВ.

## РЕЗУЛЬТАТЫ И ОБСУЖДЕНИЕ

Как уже было отмечено во введении, одной из основных проблем является недостаточно высокое кристаллическое и, вследствие этого, оптическое качество гетероструктур InAs/Si. Связано это с трудностью достижения оптимальных условий роста, позволяющих сочетать низкие температуры (ниже 450 °C) при МПЭ росте квантовых точек InAs и относительно высокие температуры при росте покровного слоя. Осуществление низкотемпературного заращивания кремнием позволяет подавить процессы сегрегации и диффузии атомов индия, но в то же время может приводить к переходу от кристаллического роста к аморфному при достижении критической толщины слоя (более 15–20 нм) [19]. В свою очередь, высокотемпературное заращивание способствует эпитаксиальному послойному росту кремния и формированию однородного кристаллического покровного слоя, однако может приводить к термическому разложению островков InAs [20]. Следует отметить, что влияние температуры роста покровного слоя и термического отжига на структурные свойства уже исследовалось в работах [14,21], однако ее влияние на оптические свойства гетероструктур InAs/Si не было продемонстрировано. Поэтому в настоящей работе было решено прежде всего исследовать влияние метода и температурного режима формирования покровного слоя на оптические свойства гетероструктур с КТ InAs. Для этого была выращена серия образцов с различными режимами осаждения кремния во время формирования покровного слоя, при этом параметры роста InAs островков оставались неизменными.

Спектры низкотемпературной ФЛ полученных образцов представлены на рисунке 1. Полученные образцы с 0,7 МС КТ InAs демонстрировали ФЛ с максимумами в диапазоне 1650–1660 нм, что хорошо согласуется с ранними работами [17]. Полная ширина на полувысоте (FWHM) для всех образцов составляла около 105±3 мэВ, что свидетельствует о формировании массивов КТ InAs, обладающих одинаковой формой и близкими размерами. В то же время известно, что существует корреляция между шириной прямой запрещенной зоны КТ InAs и их размерами. Согласно оценкам, основанным на результатах расчетов в работе [22], излучение на длинах волн 1650–1670 нм соответствует диаметрам КТ около 4–5 нм. Как будет показано далее, для полученных образцов размеры КТ InAs, внедренных в матрицу кремния, также лежали в диапазоне 4–5 нм (рисунок 4).

Анализ полученных спектров показал, что наименьшую интенсивность демонстрировал образец, при росте которого заращивание кремнием осуществлялось при температуре подложки, равной 400 °C (рисунок 1). В то же время повышением температуры заращивания в пределах 450 °C привело к увеличению интегральной интенсивности ФЛ, которое также сопровождалось небольшим красным смещением длины волны излучения. Данный факт мог быть обусловлен переходом от островкового роста кремния, происходящего при температурах порядка 250–400 °C [20], к послойному ступенчатому росту кремния. Это, в свою очередь, позволило осуществить эпитаксиальное заращивание островков InAs кремнием [23].

Как уже было сказано ранее, подъем температуры подложки выше 450 °C может приводить к термическому разложению островков InAs. Поэтому, для дальнейшего улучшения кристаллического качества гетероструктур, было решено использовать многостадийный способ заращивания кремнием, предложенный в работе [21]. Данный подход заключался в заращивании InAs наноостровков тонким пассивирующим слоем кремния без изменения температуры подложки с последующим формированием покровного слоя кремния при более высоких температурах. Ожидалось, что использование тонкого пассивирующего слоя предотвратит термическое разложение InAs в процессе повышения температуры подложки. Поэтому для образца (3) сначала в процессе повышения температуры подложки с 400 до 440 °C был сформирован слой Si толщиной 10 нм. После этого осаждение Si приостанавливалось и возобновлялось после нагрева подложки до 540 °C. В результате проведенных исследований оптических свойств было установлено, что выращенный образец демонстрировал сигнал ФЛ, в 4 раза превосходящий по интенсивности сигнал от гетероструктуры с низкотемпературным (400 °C) покровным слоем (рисунок 1, спектры (3) и (1) соответственно) и красное смещение пика ФЛ до 1670 нм.

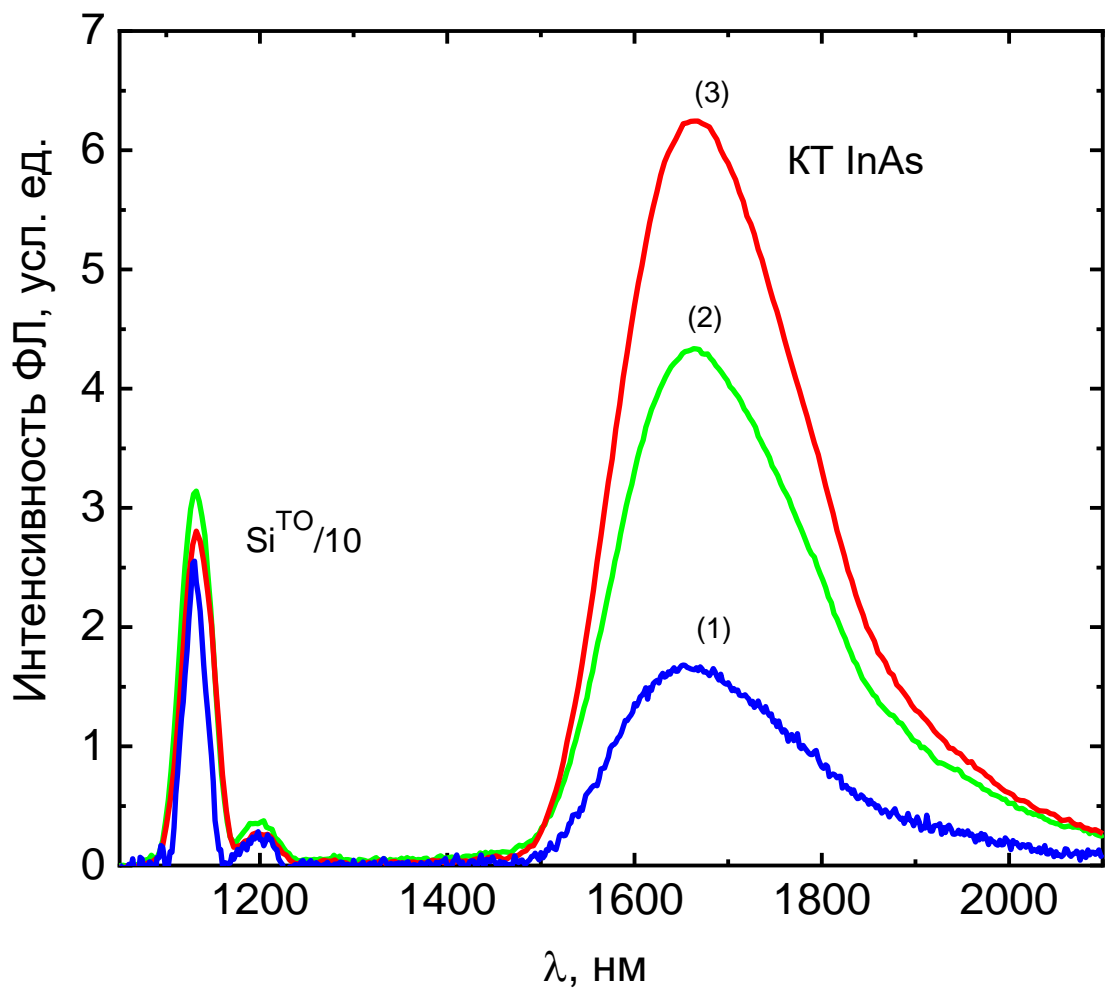

**Рис. 1.** Типичные спектры интенсивности фотолюминесценции при 10 К образцов, выращенных при различных условиях формирования покровного слоя: одностадийное заращивание при (1) 400 °C и (2) 420–450 °C; двухстадийное заращивание кремнием при 420–440 °C и 540 °C.

Более детальные исследования оптических свойств образца (3) показали, что полоса излучения КТ InAs с положением максимума на длине волны 1670 нм имеет FWHM около 105 мэВ в диапазоне температур 10–70 К. Последующее повышение температуры до 120 К приводило к линейному увеличению FWHM. На рисунке 2 представлены зависимости интегральной интенсивности фотолюминесценции (I) от плотности мощности накачки (P) в диапазоне 0,3–30 Вт/см$^2$ при температуре 10 К для образца (3) и для буферного слоя Si, выращенного на подложке кремния. Экспериментальные данные хорошо согласуются со степенным законом I пропорционально P$^B$, где показатель B связан с механизмами рекомбинации [24]. При малых интенсивностях возбуждения интенсивность ФЛ внедренных в матрицу кремния КТ InAs в несколько раз больше, чем у подложки, что указывает на большое сечение захвата фотонов возбуждения в КТ InAs. При этом для кремниевой подложки с осажденным буферным слоем рост интенсивности ФЛ был практически линейным, в то время как для КТ InAs наблюдался сублинейный (log–log) с параметром B равным 0,5 рост интенсивности ФЛ. Такой характер зависимости может быть обусловлен доминированием процессов Оже–рекомбинации при повышении мощности накачки [25,26].

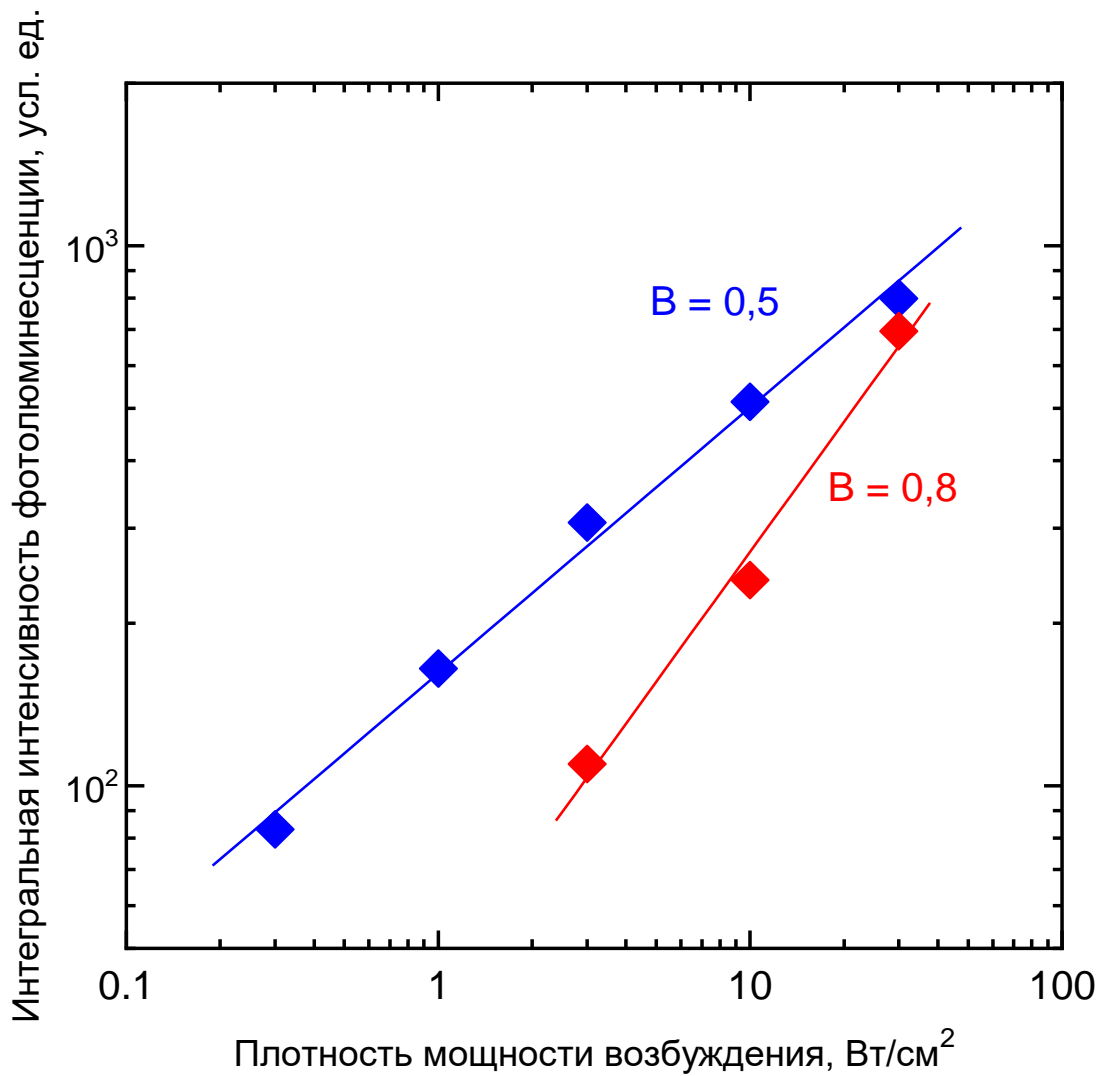

**Рис. 2.** Экспериментальные зависимости интенсивности люминесценции от интенсивности возбуждения для внедренных в слой кремния КТ InAs (синий цвет) и буферного слоя Si (красный цвет) на кремнии. Сплошные линии – аппроксимация линейной функцией.

На рисунке 3 показана зависимость интегральной интенсивности ФЛ от КТ InAs, внедренных в матрицу кремния, от обратной температуры. Интегральная интенсивность фотолюминесценции была практически постоянна при температурах ниже 50 К и снижалась при повышении температуры вплоть до 120 К. Термический спад интегральной интенсивности ФЛ сопровождался слабым красным смещением пиковой длины волны из-за зависящего от температуры сокращения запрещенной зоны [17]. Кроме того, из температурной зависимости интенсивности ФЛ можно получить величину энергии термической активации электронов, удерживаемых в потенциальной яме КТ, путем подгонки ее к уравнению Аррениуса (1), где A(T) – интегральная интенсивность ФЛ при температуре T, $A_{max}$ – интегральная интенсивность ФЛ при низких температурах; постоянная C – отношение вероятностей безызлучательных и излучательных переходов; $E_A$ – энергия активации тушения ФЛ; k — постоянная Больцмана [27]:

$$A(T) = \frac{A_{max}}{1 + C \exp\left(-\frac{E_A}{kT}\right)} \qquad (1)$$

Полученная энергия активации термического тушения ФЛ $E_A$ составила 30 мэВ и оказалась сравнима с термической энергией электронов при комнатной температуре (25 мэВ).

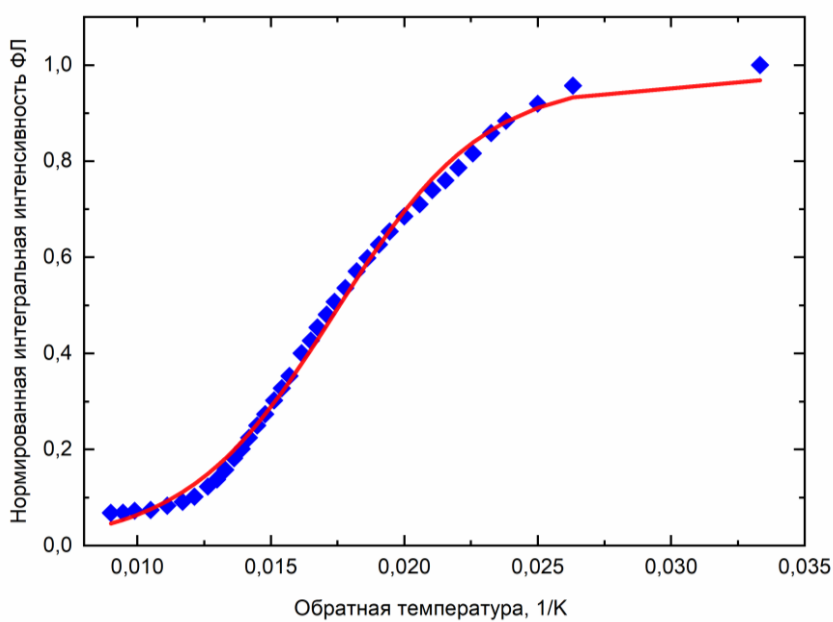

**Рис. 3.** Температурная зависимость интегральной интенсивности ФЛ для образца (3). Сплошная линия – аппроксимация по формуле Аррениуса (1).

Исследование структурных свойств образца (3) было выполнено методом просвечивающей электронной микроскопии (рисунок 4). Анализ ПЭМ–изображений показал, что после заращивания в слое кремния формируются КТ пирамидальной формы с диаметром основания около 4–5 нм и высотой около 3–4 нм. На рисунке 4а показана характерная одиночная КТ, полностью внедренная в матрицу кремния. В то же самое время были выявлены КТ, на гетерогранице которых с кремнием наблюдалось формирование дислокаций. По-видимому, они образовались на начальном этапе осаждения кремния у подножия КТ, а затем распространились вдоль боковых граней квантовых точек в процессе эпитаксиального заращивания. Вероятно тушение ….. Таким образом, температурная и мощностная зависимости в совокупности указывают на то, что вероятный механизм тушения ФЛ в КТ InAs мог быть связан с увеличением скорости безызлучательной рекомбинации на дефектах гетерограниц КТ InAs и кремния и высокой вероятностью Оже–рекомбинации при низких температурах [27].

Тем не менее, причины возникновения дислокаций такого рода остаются не до конца понятными и требуют дальнейших исследований.

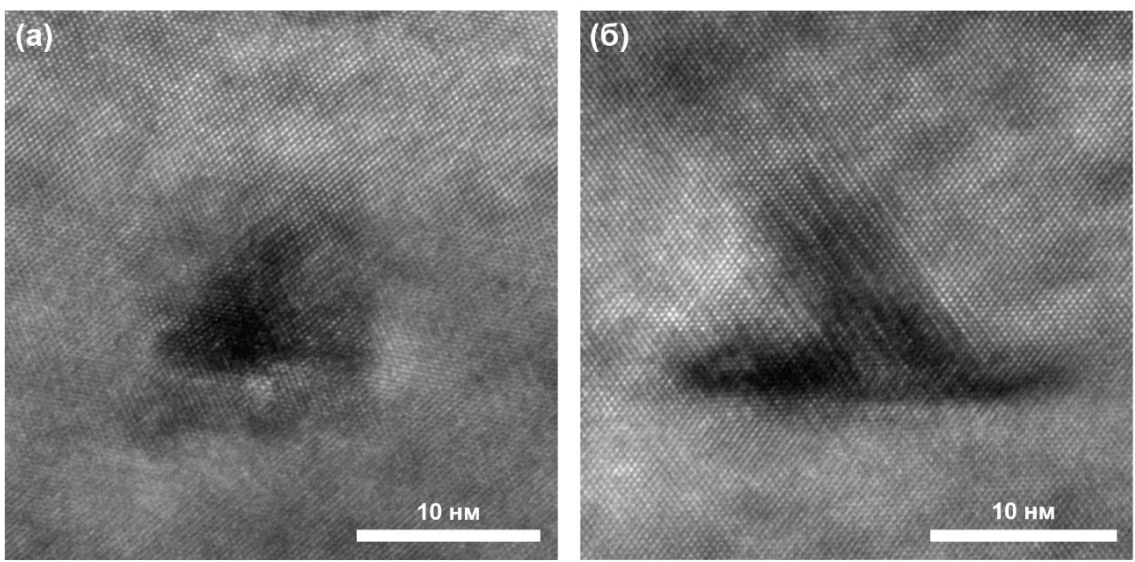

**Рис. 4.** (а) ПЭМ–изображение поперечного сечения квантовых точек InAs, внедренных в матрицу кремния, на котором видно дислокации несоответствия на границе раздела InAs/Si; (б) ПЭМ–изображение поперечного сечения одиночной бездефектной квантовой точки InAs, внедренной в матрицу кремния.

## ВЫВОДЫ

В результате выполненных работ было проведено исследование влияния режимов роста покровного слоя Si на оптические свойства гетероструктур с субмонослойными квантовыми точками InAs, внедренными в матрицу кремния. Было показано, что низкотемпературное (400–450 ℃) заращивание кремнием позволяет получить гетероструктуры, демонстрирующие фотолюминесценции в области 1,65 мкм при пониженных температурах вплоть до 120 K. В то же время было установлено, что применение двухстадийного способа заращивания InAs наноостровков кремнием позволяет повысить интенсивность фотолюминесценции за счет улучшения кристаллографического качества матрицы Si. Были проведены исследования оптических свойств полученных гетероструктур, при этом мощностная и температурная зависимости интенсивности фотолюминесценции позволили установить вероятный механизм тушения ФЛ в КТ InAs, взаимосвязанный с Оже–рекомбинацией и безызлучательной рекомбинации на дефектах гетерограниц КТ InAs и кремния. Таким образом, при дальнейшем развитии технологии формирования гетероструктур с субмонослойными КТ InAs, данные нанообъекты могут послужить основой для создания фотоприемников в ближнем ИК–диапазоне на основе кремниевых технологий.

## СПИСОК ИСТОЧНИКОВ

**ПОДПИСИ К РИСУНКАМ**

**Рис. 1.** Спектры интенсивности фотолюминесценции образцов 1 (синий цвет), 2 (зеленый цвет) и 3 (красный цвет) при 10 K.

**Fig. 1.** Photoluminescence intensity spectra of samples 1 (blue), 2 (green) and 3 (red) at 10 K.

**Рис. 2.** Температурная зависимость интегральной интенсивности фотолюминесценции от внедренных в слой кремния квантовых точек InAs. Сплошная линия – аппроксимация.

**Fig. 2.** The temperature dependence of the integral photoluminescence intensity on the InAs quantum dots embedded in Si matrix. Solid line – approximation.

**Рис. 3.** Экспериментальные зависимости интенсивности люминесценции от интенсивности возбуждения для внедренных в слой кремния КТ InAs (синий цвет) и буферного слоя Si (красный цвет) на кремнии. Сплошные линии – аппроксимация.

**Fig. 3.** Experimental dependence of luminescence intensity from excitation intensity on the InAs quantum dots embedded in Si (blue), and Si buffer layer (red) on silicon. Solid lines - approximation.

**АВТОРЫ**

Вера Вадимовна Лендяшова – инженер–исследователь, Санкт–Петербургский государственный университет, 199034, Санкт–Петербург, Россия; лаборант, Академический университет им. Ж.И. Алфёрова, 194021, Санкт–Петербург, Россия; https://orcid.org/0000-0001-8192-7614, erilerican@gmail.com

Vera Vadimovna Lendyashova – research engineer, St. Petersburg State University, 199034, Saint Petersburg, Russia; laboratory assistant, Alferov University, 194021, Saint Petersburg, Russia; https://orcid.org/0000-0001-8192-7614, erilerican@gmail.com

Игорь Владимиович Илькив – старший научный сотрудник, Санкт–Петербургский государственный университет, 199034, Санкт–Петербург, Россия; научный сотрудник, Академический университет им. Ж.И. Алфёрова, 194021, Санкт–Петербург, Россия; https://orcid.org/0000-0001-8968-3626, fiskerr@ymail.com

Igor Vladimirovich Ilkiv – senior researcher, St. Petersburg State University, 199034, Saint Petersburg, Russia; researcher, Alferov University, 194021, Saint Petersburg, Russia; https://orcid.org/0000-0001-8968-3626, fiskerr@ymail.com

Вадим Геннадьевич Талалаев – старший научный сотрудник, Санкт–Петербургский государственный университет, 199034, Санкт–Петербург, Россия; https://orcid.org/0000-0003-2559-889X, vadimtal2212@gmail.com

Vadim Gennadievich Talalaev – senior researcher, St. Petersburg State University, 199034, Saint Petersburg, Russia; https://orcid.org/0000-0003-2559-889X, vadimtal2212@gmail.com

Талгат Шугабаев – лаборант–исследователь, Санкт–Петербургский государственный университет, 199034, Санкт–Петербург, Россия; аспирант, Академический университет им. Ж.И. Алфёрова, 194021, Санкт–Петербург, Россия; https://orcid.org/0000-0002-4110-1647, talgashugabaev@mail.ru

Talgat Shugabaev – research laboratory assistant, St. Petersburg State University, 199034, Saint Petersburg, Russia; graduate student, Alferov University, 194021, Saint Petersburg, Russia; https://orcid.org/0000-0002-4110-1647, talgashugabaev@mail.ru

Родион Романович Резник – заведующий лабораторией, Санкт–Петербургский государственный университет, 199034, Санкт–Петербург, Россия; старший научный сотрудник, Академический университет им. Ж.И. Алфёрова, 194021, Санкт–Петербург, Россия; https://orcid.org/0000-0003-1420-7515, moment92@mail.ru

Rodion Romanovich Reznik – head of laboratory, St. Petersburg State University, 199034, Saint Petersburg, Russia; senior researcher, Alferov University, 194021, Saint Petersburg, Russia; https://orcid.org/0000-0003-1420-7515, moment92@mail.ru

***Георгий Эрнстович Цырлин*** – ведущий научный сотрудник, Санкт–Петербургский государственный университет, 199034, Санкт–Петербург, Россия; заведующий лабораторией, Академический университет им. Ж.И. Алфёрова, 194021, Санкт–Петербург, Россия; главный научный сотрудник, Институт аналитического приборостроения Российской академии наук, 198095, Санкт-Петербург, Россия; https://orcid.org/0000-0003-0476-3630, cirlin.beam@mail.ioffe.ru

***George Ernstovich Cirlin*** – leading researcher, St. Petersburg State University, 199034, Saint Petersburg, Russia; head of laboratory, Alferov University, 194021, Saint Petersburg, Russia; chief researcher, Institute for Analytical Instrumentation of the Russian Academy of Sciences, 198095, Saint Petersburg, Russia; https://orcid.org/0000-0003-0476-3630, cirlin.beam@mail.ioffe.ru